\documentclass[10pt, a4paper]{article}

\usepackage[final]{lrec2026} 
\usepackage{amsmath}
\usepackage{booktabs}

\title{Quantizing Whisper: How Design Choices Affect ASR Performance}

\name{
        Arthur Söhler\textsuperscript{1}, 
        Julian Irigoyen\textsuperscript{2}, 
        Andreas Søeborg Kirkedal\textsuperscript{3}
} 

\address{\textsuperscript{1} Copenhagen Business School, Copenhagen, Denmark\\
         \textsuperscript{2} Danske Bank, Copenhagen, Denmark\\
         \textsuperscript{3} Jabra (GN Group), Copenhagen, Denmark \\
         arthursoehler@gmail.com, jiri@danskebank.dk, askirkedal@jabra.com\\}

\abstract{
Large speech recognition models like OpenAI's Whisper achieve high accuracy but are difficult to deploy in resource-constrained environments due to their high memory and computational demands. This matters for low-resource and on-device settings, where compute and memory constraints often limit the practical use and evaluation of ASR systems. To address this, we present a unified, cross-library evaluation of post-training quantization (PTQ) on Whisper-small, comparing supported configurations across quantization scheme, method, granularity, and bit-width. Our study is based on four libraries—\textit{PyTorch}, \textit{Optimum-Quanto}, \textit{HQQ}, and \textit{bitsandbytes}. Experiments on \textit{LibriSpeech} \textit{test-clean} and \textit{test-other} show that dynamic \textit{int8} quantization with \textit{Optimum-Quanto} offers the best trade-off, reducing model size by 57\% while lowering Word Error Rate below the baseline. Additional experiments on Whisper-base and Whisper-tiny confirm these trends, though with more pronounced degradation at lower bit-widths. Static quantization performed worse, likely due to the absence of efficient low-bit implementations for operations such as LayerNorm and Softmax. More aggressive formats (e.g., \textit{nf4}, \textit{int3}) achieved up to 71\% compression at the cost of accuracy in acoustically challenging conditions. Our results demonstrate that carefully chosen PTQ methods can substantially reduce model size and inference cost without retraining, enabling efficient deployment of Whisper on constrained hardware. 
 \\ \newline \Keywords{automatic speech recognition, neural network quantization, model compression} }

\begin{document}

\maketitleabstract

\section{Introduction}

Automatic speech recognition (ASR) has advanced rapidly with large-scale Transformer models \cite{vaswani2017attention} such as the Whisper family \cite{radford2022robustspeechrecognitionlargescale}, which deliver state-of-the-art transcription accuracy across diverse languages and domains. However, this accuracy comes at a cost: models with hundreds of millions of parameters are difficult to deploy on edge devices, embedded systems, or latency-sensitive applications \cite{gholami2021survey, wei2024}. This challenge is especially pronounced in low-resource and on-device settings, where limited compute budgets and strict memory constraints can prevent both deployment and systematic evaluation. Post-training compression methods such as post-training quantization (PTQ) \cite{gholami2021survey, nagel2021white} therefore offer a practical path to making strong ASR models usable in settings where retraining and specialized hardware are not feasible.

Model compression techniques address this challenge by reducing computational cost in deep neural networks \cite{vanhoucke2011improving}. Among them, \emph{quantization} has emerged as one of the most practical and hardware-friendly approaches. By mapping 32-bit floating point weights and activations to lower-precision formats, quantization can shrink memory footprint and accelerate inference \cite{gray1998quantization}. The choice of method---such as PTQ, which requires no retraining, or quantization-aware training (QAT), which incorporates quantization effects during training---determines the balance between efficiency and accuracy. While QAT often yields superior results under aggressive compression, PTQ is attractive for its speed and ease of deployment \cite{wu2020integer}.

Recent years have seen rapid progress in low-bit quantization, including sub-8-bit formats (e.g., \textit{int4}, \textit{nf4}) and mixed-precision strategies designed to handle outlier activations \cite{dettmers2022llmint88bitmatrixmultiplication,shen2023efficient}. These methods have been evaluated extensively in computer vision and large language models, but their impact on ASR---and particularly on large-scale architectures like Whisper---remains underexplored.

In this paper, we take Whisper-small as a case study to evaluate post-training quantization across multiple libraries, methods, schemes, and bit-widths. We present a comparative study of Whisper PTQ in which all configurations are evaluated under a unified measurement protocol on both CPU and GPU, highlighting practical trade-offs between latency, accuracy, and compression across devices and tools. Our goal is to uncover how specific PTQ design choices translate into deployment behavior (efficiency and robustness) in realistic scenarios, where library support and overheads can materially affect outcomes. We additionally confirm our findings across model sizes by performing experiments on Whisper-base and Whisper-tiny, the two smallest variants of the Whisper family.

Our contributions are:

\textbf{(1) Unified evaluation framework:} Systematic cross-library comparison of PTQ for Whisper under a controlled, unified evaluation protocol, enabling direct comparison of library-specific trade-offs.

\textbf{(2) Comparative analysis of PTQ design choices:} We compare the effects of quantization scheme (dynamic vs. static), method (symmetric vs. asymmetric), granularity (per-tensor vs. per-channel), and bit-width (\textit{int8} to \textit{int3}/\textit{nf4}) across representative libraries.

\textbf{(3) Device-specific deployment insights:} Parallel evaluation on CPU and GPU revealing different optimal configurations per device and acoustic condition (clean vs. acoustically challenging).

\textbf{(4) Cross-model validation:} Verification that the main findings hold across Whisper-small, Whisper-base, and Whisper-tiny, demonstrating robustness of deployment conclusions across model sizes.

\section{Related Work}

While its theoretical foundations date back decades~\cite{gray1998quantization}, modern neural network quantization has evolved rapidly, with several surveys providing systematic overviews~\cite{gholami2021survey, nagel2021white, wei2024}. These works classify methods into post-training quantization and quantization-aware training, and outline factors such as bit-width, quantization granularity, and calibration strategies that strongly influence performance.

In Transformer-based architectures, \citet{kim2021i-bert} proposed I-BERT, an integer-only Transformer that approximates nonlinear operations for complete deployment on integer hardware. \citet{kim2022integeronlyzeroshotquantizationefficient} further introduced a zero-shot static quantization approach for ASR models such as Conformer and QuartzNet, achieving $4\times$ compression without real training data by synthesizing calibration inputs. \citet{wu2020integer} demonstrated that with per-channel weight quantization and entropy-based activation calibration, PTQ can maintain accuracy within 1\% of full precision.

Recent efforts have focused on stabilizing quantization for large Transformer models by handling outlier activations. \citet{dettmers2022llmint88bitmatrixmultiplication} introduced LLM.int8(), a mixed-precision method that preserves high precision for outlier dimensions, enabling robust 8-bit inference for models up to 175B parameters. \citet{shen2023efficient} explored low-precision floating-point formats such as \textit{fp8}, reporting improved accuracy and flexibility compared to \textit{int8} in outlier-heavy models.

Prior work has also examined Whisper quantization in specific toolchains and limited method sets. For example, \citet{andreyev2025quantizationopenaiswhispermodels} evaluates different bit-widths using \textit{whispercpp} on \textit{LibriSpeech} corpus \citeplanguageresource{LibriSpeech} and reports latency and accuracy trade-offs for edge deployment scenarios. In contrast, our work focuses on a controlled, cross-library comparison spanning multiple PTQ libraries and design choices (scheme, granularity, and bit-width) under one measurement protocol across CPU and GPU.

More recent work has explored low-bit quantization specifically for speech and audio models. \citet{feng2025edgeasr} propose Edge-ASR, a comprehensive benchmark of post-training quantization methods applied to Whisper and other edge-ASR models. \citet{kang-kim-2025-genptq} propose GenPTQ, a mixed-precision post-training quantization method that performs efficient layer-wise bit allocation based on gradient-driven sensitivity analysis, achieving strong compression with minimal WER degradation. Beyond speech recognition, \citet{khandelwal2025ptqdit} study post-training quantization for audio diffusion transformers, highlighting challenges related to activation variability and outliers in transformer-based audio models. These works primarily focus on model-specific optimization strategies, whereas our study provides a cross-library evaluation of quantization approaches under a unified experimental framework.

While most prior work has focused on vision and text models, the techniques—especially sub-8-bit quantization, mixed-precision schemes, and outlier handling—are directly relevant to ASR. However, to our knowledge, there is no comparative evaluation of quantization libraries that jointly contrasts (i) activation method (dynamic vs. static), (ii) granularity (per-channel vs. per-tensor), (iii) weight-only vs. weight+activation quantization, and (iv) multiple bit-widths, while evaluating accuracy, latency, and compression on both CPU and GPU under a single protocol. This work addresses that gap by assessing PTQ strategies and their trade-offs for Whisper-small, Whisper-base, and Whisper-tiny.

\section{Background}

Quantization reduces the memory and computational cost of neural networks by mapping high-precision values (typically 32-bit floating point) to lower-precision representations. In practice, this means storing weights and activations in compact formats such as \textit{int8} or \textit{fp8}, with memory usage scaling approximately as $b/32$ relative to \textit{fp32}, yielding a compression ratio of $32/b$ where $b$ is the bit-width \citep{gholami2021survey,nagel2021white}.

Formally, a quantizer can be written as
\begin{equation}
Q(x) = \text{clip}\Big(\text{round}\big(\tfrac{x}{s}\big), q_{\min}, q_{\max}\Big),
\end{equation}
where $s$ is a scaling factor, and $[q_{\min}, q_{\max}]$ defines the representable integer range for a given bit-width. The effectiveness of this mapping depends not only on the chosen bit-width, but also on how scales are applied across the network. 

This is captured by the notion of \textit{granularity}. In \textit{per-tensor} quantization, a single scale is shared across an entire tensor, which is efficient but may distort values when distributions vary widely. \textit{Per-channel} quantization instead assigns a scale to each output channel, better capturing local statistics and improving accuracy in deep Transformers, though at higher storage and compute cost \citep{gholami2021survey}. A middle ground is \textit{per-group} quantization, which clusters channels into groups with shared scales to balance efficiency and accuracy \citep{yao2022zero}. 

Beyond granularity, quantization can be applied either post-training (PTQ), which avoids retraining but may reduce accuracy at low bit-widths, or during training (QAT), which improves robustness at the cost of additional training \citep{han2015learningweightsconnectionsefficient, gholami2021survey}. 

Additionally, quantization methods differ on how they distribute values within the quantized range. Symmetric quantization centers values around zero, whereas asymmetric quantization shifts the zero-point to better match non-zero-centered distributions of weights \citep{nagel2021white}. 

Finally, schemes differ in how scaling factors and zero-points are determined. \textit{Static quantization} fixes them in advance using calibration data, while \textit{dynamic quantization} computes them at runtime. 

These design choices become especially important for Transformer-based ASR models such as Whisper-small. Operations like \textit{LayerNorm}, \textit{Softmax}, and \textit{GELU} are highly sensitive to reduced precision, and activations often exhibit heavy-tailed distributions \citep{dettmers2022llmint88bitmatrixmultiplication,shen2023efficient}. As a result, quantization can compromise robustness, underscoring the need for careful selection of quantization strategies in deployment.

\section{Data}

We conduct our experiments on the English-language subsets \textit{test-clean} and \textit{test-other} of the \textit{LibriSpeech} corpus \citeplanguageresource{LibriSpeech}. Dataset statistics are shown in Table 1.

\begin{table}[t]
\renewcommand{\arraystretch}{1.3}
\centering
\label{tab:data}
\scalebox{0.7}{\begin{tabular}{lccccc}
\toprule
\textbf{Subset} & \textbf{Hours} & \textbf{Minutes/} & \textbf{Female} & \textbf{Male} & \textbf{Total} \\
 &  & \textbf{speaker} & \textbf{speakers} & \textbf{speakers} & \textbf{ speakers} \\
\midrule
test-clean & 5.4 & 8 & 20 & 20 & 40  \\
test-other & 5.1 & 10 & 17 & 16 & 33 \\
\bottomrule
\end{tabular}}
\caption{Dataset statistics for \textit{test-clean} and \textit{test-other}.}
\end{table}

These test sets have been used to evaluate ASR systems for many years. \textit{test-clean} represents an easy evaluation task, whereas \textit{test-other} is a more challenging dataset. The original \textit{LibriSpeech} data consists of read-aloud books and was divided into a \textit{clean} and an \textit{other} partition based on Word Error Rate (WER) scores produced by a hybrid ASR system with an acoustic model trained on a subset of the Wall Street Journal corpus and a bigram LM estimated from the text of the respective books. The speakers were ranked according to WER and the data partitioned into two sets of roughly equal sizes. \textit{test-clean} consists of randomly selected speakers from the \textit{clean} partition. \textit{test-other} specifically consists of speakers from the 3rd quartile according to WER. This sampling is intended to create a more challenging test dataset. Gender balance is ensured at the speaker level~\cite{panayotov2015librispeech}.

\section{Methods}

We evaluate PTQ on Whisper-small, a 244M-parameter Transformer-based ASR model pre-trained for multilingual and multitask speech recognition \cite{radford2022robustspeechrecognitionlargescale}. 

Depending on library support, we apply quantization across a range of bit-widths (\textit{int8}, \textit{int4}, \textit{int3}, \textit{nf4}, and \textit{fp8}) and compare four widely used PTQ libraries: \textit{PyTorch}, offering native dynamic \textit{int8} quantization on CPU; \textit{Optimum-Quanto} (hereafter \textit{Quanto}), supporting both integer and low-precision floating formats across CPU, GPU, and MPS backends; \textit{HQQ}, a quantization library based on half-quadratic optimization with configurable group sizes; and \textit{bitsandbytes (BNB)}, which provides GPU-only implementations of normalized formats such as \textit{nf4} with optional double quantization. Together, these libraries cover a diverse spectrum of formats, methods, and quantization schemes, making them representative of the practical choices available to users when designing model compression pipelines.

Performance is evaluated along three dimensions. Accuracy is measured using WER and Character Error Rate (CER). Efficiency is captured through the Real-Time Factor (RTF), quantifying inference speed relative to audio duration. Finally, we report model size reduction relative to the full-precision \textit{fp32} baseline. To ensure comparability, all configurations were evaluated under the same preprocessing pipeline, dataset splits, batch size, decoding procedure, hardware selection, and timing protocol. RTF was computed as total timed generation time divided by total audio duration over the evaluation set. Because all measurements were collected on the HPC infrastructure described in Section~\ref{sec:Acknowledgements}, the reported RTF values should be interpreted primarily as relative comparisons across configurations under a fixed setup, rather than as direct estimates of edge-device latency. Inference time was measured only around the model generation step, with CUDA synchronization on GPU and warm-up runs before timed evaluation. Audio inputs were converted to model input features using the standard WhisperProcessor from the Transformers library. In the adopted Transformers implementation, Whisper used its default fixed 30\,s front end; accordingly, shorter utterances were zero-padded to the 30\,s window. The reported RTF values therefore reflect this fixed-window inference setup rather than variable-length audio processing. Relative RTF differences nevertheless remain directly comparable across quantization settings because all configurations were evaluated under the same protocol.

For static quantization, calibration used a randomly selected 10\% subset of the full evaluation data, processed with the same feature-extraction pipeline as the test data.

While the main study focuses on Whisper-small, we also run a limited follow-up study on Whisper-tiny and Whisper-base under the same dataset, settings, and measurement procedure. The cross-model follow-up is intended as a validation of the main deployment trends, not as a second exhaustive benchmark. We therefore restrict Whisper-tiny and Whisper-base to the strongest dynamic configurations from the Whisper-small study. Static quantization is omitted in this follow-up because it underperformed on Whisper-small and would substantially increase experimental volume without changing the main practical conclusion.

\section{Results}

\subsection{Quantizing Whisper-small}
Table 2 summarizes the best-performing quantized models relative to the full-precision baseline.

\begin{table}[t]
\renewcommand{\arraystretch}{1.3}
\centering
\label{tab:results}
\scalebox{0.62}{\begin{tabular}{lcccccc}
\toprule
\textbf{Device/Method} & \textbf{WER$_c$} & \textbf{WER$_o$} & \textbf{CER$_c$} & \textbf{CER$_o$} & \textbf{RTF} & \textbf{Size Red.} \\
\midrule
\multicolumn{7}{c}{\textbf{CPU}} \\
\midrule
Baseline (fp32) & 3.48 & 11.88 & 1.02 & 3.62 & 0.121 & -- \\
PyTorch int8 (dyn.) & 3.72 & 13.67 & 1.11 & 4.21 & 0.077 & 57\% \\
HQQ int4 (dyn.) & 3.52 & 14.09 & 1.09 & 4.38 & 0.155 & 69\% \\
Quanto int8/fp8 (stat.) & 5.95 & 15.92 & 1.83 & 5.03 & 0.169 & 57\% \\
\midrule
\multicolumn{7}{c}{\textbf{GPU}} \\
\midrule
Baseline (fp32) & 3.48 & 11.88 & 1.02 & 3.62 & 0.006 & -- \\
Quanto int8 (dyn.) & 3.41 & 10.65 & 0.97 & 3.29 & 0.008 & 57\% \\
BNB nf4 (dyn.) & 3.54 & 13.49 & 1.05 & 4.05 & 0.008 & 70\% \\
HQQ int3 (dyn.) & 4.11 & 12.93 & 1.22 & 3.77 & 0.019 & 71\% \\
\bottomrule
\end{tabular}}
\caption{Selected best-performing dynamic (dyn.) and static (stat.) quantization configurations for Whisper-small on \textit{LibriSpeech} \textit{test-clean} (c) and \textit{test-other} (o).}
\end{table}

On \textbf{CPU}, \textit{PyTorch} dynamic \textit{int8} delivered the fastest inference (RTF 0.077; 36.4\% faster than the 0.121 baseline) with only a small accuracy drop. \textit{HQQ} dynamic \textit{int4} preserved accuracy on clean speech while achieving the largest compression (69\%). In contrast, static \textit{int8}/\textit{fp8} quantization with \textit{Quanto} degraded performance substantially, confirming that Whisper’s architecture is ill-suited to static quantization.

On \textbf{GPU}, \textit{Quanto} dynamic \textit{int8} not only matched but slightly outperformed the baseline on the more challenging \textit{test-other} split (WER$_{\text{other}}$ = 10.65, CER$_{\text{other}}$ = 3.29), while reducing model size by 57\%. \textit{BNB} \textit{nf4} offered a 70\% size reduction with minimal accuracy loss on \textit{test-clean}, but suffered on \textit{test-other}. \textit{HQQ} \textit{int3} achieved the smallest size (71\% reduction) but at the cost of higher error rates.

Overall, dynamic quantization consistently outperformed static quantization in both accuracy and speed. \textit{int8} proved the most reliable setting across devices, while more aggressive formats (\textit{nf4}, \textit{int3}) enabled extreme compression but compromised robustness, particularly under acoustically challenging conditions.

\subsection{Scaling Across Whisper Model Sizes}
To assess whether the main conclusions are specific to Whisper-small or stable within the Whisper model family, we evaluate Whisper-tiny and Whisper-base on \textit{LibriSpeech} splits using the strongest quantization configurations from the Whisper-small study. Table 3 shows that the overall trends remain consistent across model sizes: dynamic \textit{int8} remains the most robust operating point, while more aggressive formats trade robustness (especially on \textit{test-other}) for additional compression.

\begin{table}[t]
\label{tab:base-tiny-quantization-new}
\renewcommand{\arraystretch}{1.3}
\centering
\scalebox{0.63}{\begin{tabular}{lcccccc}
\toprule
\textbf{Device/Method} & \textbf{WER$_c$} & \textbf{WER$_o$} & \textbf{CER$_c$} & \textbf{CER$_o$} & \textbf{RTF} & \textbf{Size Red.} \\
\midrule
\multicolumn{7}{c}{\textbf{Whisper-base}} \\
\midrule
Baseline (fp32)    & 5.08  & 12.87 & 1.93 & 6.76  & 0.0022 & -- \\
Quanto int8 (dyn.) & 5.10  & 14.72 & 1.93 & 7.60  & 0.0037 & 36.2\% \\
BNB nf4 (dyn.)     & 6.05  & 18.09 & 2.31 & 10.16 & 0.0036 & 52.1\% \\
\midrule
\multicolumn{7}{c}{\textbf{Whisper-tiny}} \\
\midrule
Baseline (fp32)    & 7.60  & 23.69 & 2.98 & 12.84 & 0.0015 & -- \\
Quanto int8 (dyn.) & 7.64  & 24.64 & 2.99 & 13.24 & 0.0025 & 19.3\% \\
BNB nf4 (dyn.)     & 11.16 & 32.19 & 4.78 & 20.20 & 0.0026 & 37.5\% \\
\bottomrule
\end{tabular}}
\caption{Quantization results for Whisper-base and Whisper-tiny using the strongest dynamic (dyn.) configurations on \textit{LibriSpeech} \textit{test-clean} (c) and \textit{test-other} (o).}
\end{table}

\section{Discussion}

\subsection{Trade-offs Between Different Quantization Methods}
On CPUs, \textit{PyTorch} dynamic \textit{int8} consistently achieved the fastest inference. Its advantage likely stems from using a per-tensor asymmetric scheme, which applies a single scale across an entire tensor. This approach simplifies computation and reduces the overhead of quantization and dequantization, explaining the strong runtime performance. The trade-off, however, is lower representational flexibility with quantization parameters only calculated per-tensor, which contributed to weaker robustness on the acoustically diverse \textit{test-other} dataset. However, with a 57\% size reduction, it still offers clear advantages over the baseline.

On the GPU, \textit{Quanto} dynamic \textit{int8} took a different approach. It prioritizes accuracy over speed. It was slower but outperformed the \textit{fp32} baseline on WER and CER on both \textit{test-clean} and \textit{test-other}, with the most striking improvement seen under acoustically challenging conditions. One plausible explanation is its symmetric, per-channel scheme, which aligns well with Whisper’s near-zero-centered weight distributions and assigns independent scales to each channel. This granularity can better preserve fine-grained variation across channels, yielding higher accuracy but at the cost of additional computational overhead. Similar to the \textit{PyTorch} model, this model comes with a 57\% size reduction, a strong compression relative to the baseline.

Libraries enabling more aggressive formats, such as \textit{HQQ} (\textit{int4}, \textit{int3}) and \textit{BNB} (\textit{nf4}), pushed compression further—up to ~70\%—but consistently degraded robustness, especially in acoustically challenging environments. These results highlight that precision below 8-bit remains less reliable in real-world ASR applications.

Taken together, these comparisons show that library differences are largely driven by the schemes and bit-widths they implement. Per-tensor quantization favors efficiency, making \textit{PyTorch} \textit{int8} attractive when low latency is the main deployment priority. Per-channel quantization favors accuracy, as shown by \textit{Quanto} \textit{int8}, which is more suitable when robustness is critical. More aggressive low-bit formats are best reserved for deployments where extreme compression outweighs the need for reliability.

The additional results on Whisper-base and Whisper-tiny (Table 3) follow the same qualitative pattern: dynamic \textit{int8} remains the most robust operating point, while \textit{nf4} increases compression but incurs a larger error increase—especially on \textit{test-other}. This suggests that the robustness–compression trade-off observed for Whisper-small is stable across smaller Whisper variants, even though absolute WER/CER differs by model size.

\subsection{Dynamic vs. Static Quantization}
In theory, static quantization should reduce runtime overhead by fixing scales in advance, trading a small loss in accuracy for faster inference. Surprisingly, in Whisper-small we observed the opposite: static quantization was both slower and less accurate.

One possible explanation is that operations such as \textit{LayerNorm} and \textit{Softmax} lack efficient low-bit implementations, forcing repeated dequantization that cancels the expected speed gains of static quantization. On top of this, as expected, fixed calibration scales limited robustness under shifting distributions, especially in acoustically challenging conditions.

By contrast, dynamic quantization adapted better at runtime, preserving accuracy and delivering faster inference. This makes dynamic quantization the most reliable choice in our evaluation for Whisper-small, despite theoretical expectations to the contrary.

\subsection{Clean vs. Acoustically Challenging Speech}

\begin{figure}[b]
\begin{minipage}[b]{0.95\linewidth}
  \centering
  \centerline{\includegraphics[width=8.5cm]{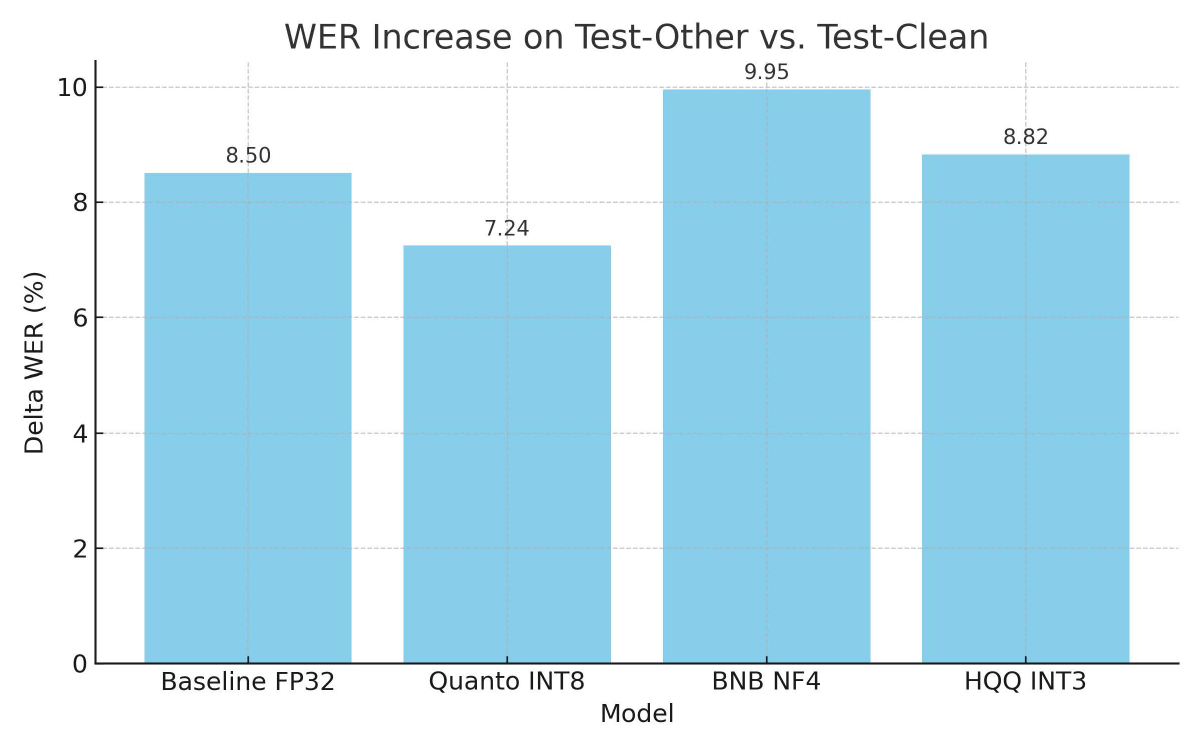}} 
\end{minipage}
\caption{WER increase on \textit{test-other} relative to \textit{test-clean} for selected quantized models of Whisper-small. Lower-bit-width configurations (\textit{nf4}, \textit{int3}) show larger deltas, highlighting the trade-off between compression and robustness.}
\label{fig:tradeoff}
\end{figure}

Across nearly all configurations, quantized models suffered larger accuracy drops on \textit{test-other} than on \textit{test-clean}, indicating reduced robustness under the more challenging conditions represented by the \textit{test-other} split. As shown in Figure \ref{fig:tradeoff}, lower-bit formats such as \textit{BNB} \textit{nf4} and \textit{HQQ} \textit{int3} showed the steepest increases in WER, underlining how aggressive compression disproportionately reduces robustness.

Interestingly, \textit{Quanto} dynamic \textit{int8} was an exception: it not only matched the baseline on \textit{test-clean} but also outperformed it on \textit{test-other}. This suggests that under certain conditions, quantization can act as a form of regularization, stabilizing predictions in acoustically challenging environments. However, this effect appears limited to moderate settings (e.g., \textit{int8}); once precision drops further, errors compound rapidly under adverse acoustic conditions. A similar robustness gap between \textit{test-clean} and \textit{test-other} is also evident for Whisper-base and Whisper-tiny, where \textit{nf4} degrades more strongly than \textit{int8} on \textit{test-other}.

For deployment, this highlights that 8-bit precision is a safe minimum in real-world scenarios, while lower-bit formats should be applied selectively and only when acoustic variability is less of a concern.

\subsection{Deployment Implications}
Our results point to clear deployment strategies for Whisper-small and its variants under different constraints. Table 4 summarizes these strategies as rule-of-thumb guidance derived from our evaluation protocol. On CPUs, \textit{PyTorch} dynamic \textit{int8} offers the best speed–accuracy trade-off, while \textit{HQQ} dynamic \textit{int4} offers stronger compression when memory is the tighter constraint. On GPUs, \textit{Quanto} dynamic \textit{int8} is the most robust choice, delivering a 57\% size reduction while preserving or even improving accuracy, including in acoustically challenging conditions.
More aggressive formats such as \textit{HQQ} \textit{int3} or \textit{BNB} \textit{nf4} are suitable only in memory-constrained scenarios where some accuracy loss is acceptable, and ideally applied selectively to less critical layers. Finally, in acoustically diverse environments, at least 8-bit precision should be maintained for sensitive components such as attention and \textit{LayerNorm} to avoid robustness degradation.
These guidelines emphasize that quantization choices must be tailored not only to model architecture, but also to deployment context, device, and acoustic variability.

\begin{table}[t]
\label{tab:quantization-model-comparison}
\centering
\scalebox{0.76}{\begin{tabular}{lll}
\toprule
\textbf{Priority} & \textbf{Best Model} & \textbf{Trade-off} \\
\midrule
\textbf{Fastest Inference} & PyTorch & Higher WER \\
 & Dynamic int8 & compared to baseline \\
\midrule
\textbf{Best Accuracy} & Quanto & Higher RTF \\
 & Dynamic int8 & compared to baseline \\
\midrule
\textbf{Best Size Reduction} & HQQ & Higher WER and RTF \\
 & Dynamic int3 & compared to baseline \\
\bottomrule
\end{tabular}}
\caption{Comparative Summary of Quantization Methods - Optimal Approaches Based on Deployment Priorities}
\end{table}

\section{Conclusion}
This study evaluated post-training quantization of Whisper-small across four libraries and multiple bit-widths, providing a controlled cross-library comparison of supported scheme, method, and granularity choices. On \textit{LibriSpeech} \textit{test-clean} and \textit{test-other}, dynamic quantization consistently outperforms static quantization in both accuracy and inference speed for this architecture. \textit{Quanto} dynamic \textit{int8} emerged as the best overall configuration for GPU deployment, achieving 57\% model-size reduction while matching baseline accuracy on \textit{test-clean} and even exceeding it on \textit{test-other}. On CPU, \textit{PyTorch} dynamic \textit{int8} delivered the fastest inference in our experiments, while \textit{HQQ} dynamic \textit{int4} offers a strong compromise when memory is limited.

From a high-level perspective, our results indicate that within this model family and library-supported configurations, dynamic quantization was more reliable than static quantization. Furthermore, the most accurate results were obtained with configurations using symmetric, per-channel quantization rather than asymmetric, per-tensor approaches, especially under acoustically variable conditions. In practice, \textit{int8} is the most reliable precision across devices; more aggressive formats (e.g., \textit{nf4}, \textit{int3}) provide larger compression but degrade robustness on \textit{test-other}. We also observe a mild regularization effect: dynamic \textit{int8} can match or even exceed \textit{fp32} on acoustically challenging speech, but this benefit does not persist below 8-bit precision. Taken together, the \emph{scheme}, \emph{method}, and \emph{granularity} materially shape the accuracy--speed--size trade-off and should be chosen to match device constraints and expected acoustic conditions. Ultimately, our findings show that thoughtful quantization design enables deployment of more advanced speech recognition models in resource-constrained environments.

\section{Limitations and Future Work}

First, this work is designed as a controlled study to compare cross-library PTQ effects under a shared evaluation protocol. Because the study is constrained by library support, not all combinations of quantization scheme, method, granularity, and bit-width could be evaluated; the comparison is therefore structured rather than exhaustive. We restrict evaluation to the \textit{LibriSpeech} \textit{test-clean} and \textit{test-other} splits, which provide a reproducible benchmark but do not capture the full diversity of domains, accents, and spontaneous speech encountered in practice. Extending the evaluation to broader domains, including multilingual data, is an important next step.

Second, we focus on PTQ to reflect scenarios where retraining is infeasible; quantization-aware training may achieve higher accuracy at more aggressive bit-widths. In addition, ONNX Runtime is excluded because it relies on a different execution stack involving model export and graph-level optimizations, introducing additional variables that make its quantization and runtime behavior not directly comparable to the native-library workflows evaluated in this study. Both represent complementary directions for future work.

Third, our analysis targets the smaller Whisper variants as representative Transformer-based ASR models. Larger variants such as Whisper-medium and Whisper-large are excluded from this study, as our focus is on smaller models that are most relevant for deployment in resource-constrained environments. Results may differ for the larger Whisper models or for other ASR architectures.

Future work will expand evaluation to additional datasets and model families, and investigate mixed-precision and layer-wise strategies that apply aggressive quantization selectively while preserving accuracy in sensitive components.

\section{Acknowledgements}\label{sec:Acknowledgements}

We thank Jabra and GN Group for supporting this research. Computational experiments were performed on the Danish e-Infrastructure Consortium (DeiC) National HPC facilities, utilizing Lenovo ThinkSystem SR675 V3 nodes equipped with dual AMD EPYC 9454 processors (2.75 GHz, 192 vCPUs total), 768 GB DDR5-4800 memory, and four NVIDIA Hopper H100-SXM5 GPUs (80 GB HBM3) per node.

\section{Bibliographical References}\label{sec:reference}

\bibliographystyle{lrec2026-natbib}
\bibliography{lrec2026-example}

\section{Language Resource References}
\label{lr:ref}
\bibliographystylelanguageresource{lrec2026-natbib}
\bibliographylanguageresource{languageresource}

\end{document}